\documentclass[twocolumn,showpacs,prl,aps,superscriptaddress,amsmath,amssymb,floatfix]{revtex4}

\usepackage{graphicx}

\begin{document}

\title{Transient time-domain resonances and the time scale for tunneling}
\author{Gast\'on Garc\'{\i}a-Calder\'on}
\email{gaston@fisica.unam.mx}
\affiliation{Instituto de F\'{\i}sica,
Universidad Nacional Aut\'onoma de M\'exico,
Apartado Postal {20 364}, 01000 M\'exico, Distrito Federal, M\'exico}
\author{Jorge Villavicencio}
\email{villavics@uabc.mx}
\affiliation{Facultad de Ciencias,
Universidad Aut\'onoma de Baja California,
Apartado Postal 1880, 22800 Ensenada, Baja California, M\'exico}

\date{\today}

\begin{abstract}
Transient {\it time-domain resonances} found recently in time-dependent solutions to Schr\"{o}dinger's equation are used to investigate the issue of the tunneling time in rectangular potential barriers. In general, a time frequency analysis shows that these transients have frequencies above the cutoff frequency associated with the barrier height, and hence correspond to non-tunneling processes. We find, however, a regime characterized by the barrier opacity, where the peak maximum $t_{max}$ of the {\it time-domain resonance} corresponds to under-the-barrier tunneling. We argue that $t_{max}$ represents the relevant tunneling time scale through the classically forbidden region. 
\end{abstract}

\pacs{03.65Xp.,0365.Ca.,73.40.Gk.}

\maketitle
Tunneling refers to the possibility that a particle traverses through a classically forbidden region. In the energy domain the solution to Schr\"odinger's equation at a fixed energy $E$ is a subject discussed in every quantum mechanics textbook. In the time domain, however, there are still aspects open to scrutiny. A problem that has remained controversial over the years is the tunneling time problem, that may be stated by the  question: How long does it take to a particle to traverse a classically forbidden region?  Different authors have proposed and defended different views in answering the above question~\cite{hauge,landauer,ghose,muga}. 

In recent work we have investigated the effect of the transient solutions to the time-dependent
Schr\"odinger's equation for cutoff wave initial conditions (quantum shutter) on the tunneling process~\cite{gcr97,gcv01,gcv02}. In particular we found that just across the tunneling barrier, the probability density as a function of time may exhibit a transient steucture that we have named {\it time-domain resonance}. The peak value $t_{max}$ of this structure represents the largest probability of finding the particle at the barrier width, $L$. More recently, in collaboration with Delgado and Muga~\cite{gcvdm02}, we considered a time-frequency analysis~\cite{mb00,cohen}, to show the existence of under-the-barrier transients (forerunners)in very broad barriers. However, these occur along a finite region of the potential, and hence do not allow to characterize the time scale associated to the tunneling process through the full classically forbidden region. On the other hand, in collaboration with Yamada~\cite{gcvy03} we have recently established the equivalence of our formulation with the notion of `passage time' in the real Feynman histories approach. The `passage time' yields the traversal time through a barrier region though it does not
distinguish between processes from above or below the barrier height. The above considerations indicate that it is not clear under what conditions the {\it time-domain resonances} found at the barrier edge $x=L$  correspond to  genuine tunneling processes. 

The aim of this work is to show the existence of a regime, characterized by the opacity of the system, where the {\it time-domain resonance} maximum, $t_{max}$, corresponds to under-the-barrier tunneling. We argue that this time scale provides the tunneling time through the classically forbidden region.  

Our approach to the tunneling time problem is based on a model that deals with an explicit solution~\cite{gcr97} to the time-dependent  Schr\"{o}dinger's equation
\begin{equation}
\left [i\hbar \frac{\partial }{\partial t} + \frac{\hbar ^{2}}{2m}\frac{\partial ^{2}}{\partial x^{2}}%
-V(x) \right ]\Psi (x,t)=0
\label{tdse}
\end{equation}
for an arbitrary potential $V(x)$, defined in the region ($0\leq x\leq L$),
that vanishes outside that region. We consider the problem 
of the time evolution of a cutoff plane wave 
%
%$\Psi (x,k;t=0)=\Theta(-x)(e^{ikx}-e^{-ikx})$
%
\begin{equation}
\Psi (x,t=0) =\left\{ 
\begin{array}{cc}
e^{ikx}-e^{-ikx}, & \quad x \le 0, \\ 
0, & \quad x>0,
\end{array}
\right.   
\label{inicon}
\end{equation}
following the instantaneous opening at $t=0$ of a quantum shutter at $x=0$. 
Along the tunneling region the solution reads,
\begin{eqnarray}
\Psi^i(x,k;t)=&&\phi_k(x)M(y_k)-\phi_{-k}(x)M(y_{-k})\nonumber\\
&&-\sum_{n=-\infty}^{\infty} \phi_n(x) M(y_{k_{n}}), \,\,\,(0 \leq x \leq L)\,\,\,\,\,
\label{psii}
\end{eqnarray}
where the $\phi_{\pm k}(x)$'s refer to the stationary solutions of the problem and 
$\phi_n(x)=2iku_n(0)u_n(x)/(k^2-k_n^2)$.
Similarly, the solution $\Psi^e(x,k;t)$ for the external or transmitted region 
($x \geq L$), is given by \cite{gcv01},

\begin{eqnarray}
\Psi^e(x,k;t)=&&T_{k}M(y_k)-T_{-k}M(y_{-k})\nonumber\\
&& -i\sum\limits_{n=-\infty }^{\infty }T_{n}M(y_{k_{n}}), \,\,\,\,\,(x \geq L) 
\label{psie}
\end{eqnarray}
where the $T_{\pm k}$'s refer to the transmission amplitudes, and the
factor $T_n=2iku_n(0)u_n(L)\exp (-ik_n L)/(k^2-k_n^2)$. In Eqs.\ 
(\ref{psii}) and (\ref{psie}) the coefficients $\chi_n$ and $T_n$ are given
in terms of the resonant eigenfunctions $\{u_n (x)\}$ with complex
energy eigenvalues $E_n=\hbar^2k_n^2/2m$, with $k_n=a_n-ib_n$
($a_n, b_n > 0$). The resonant sums in Eqs.\ (\ref{psii}) and (\ref{psie})
run over the full set of complex poles $\{k_{n}\}$. The $M^{\prime }s$ are defined as ~\cite{gcr97},
\begin{equation}
M(y_{q})=\frac{1}{2}e^{imx^{2}/2\hbar t}w(iy_{q})
\label{m}
\end{equation}
where $w$ is the complex error function~\cite{abrwtz}defined as 
$w(z)=\exp (-z^{2}){\rm erfc}(-iz)$, with arguments 
$y_{q}(x,t)=e^{-i\pi /4}(m/2\hbar t)^{1/2}[x-\hbar qt/m]$, 
where $q=\pm k,$ $k_{n}$. 

\begin{figure}
\rotatebox{0}{\includegraphics[width=3.3in]{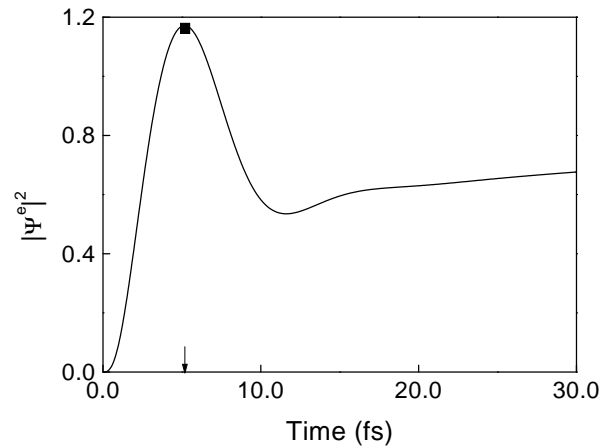}}
\caption{Time evolution of $|\Psi^e|^2$ (solid line), at $x=L=4.0$ nm.
A full square indicates the position of the maximum of the 
{\it time-domain resonance}, at $t_{max}=5.17$ fs.}
\label{fig1}
\end{figure}

We shall begin  exploring the issue of the tunneling time  scale
by first considering specific examples of rectangular potential barriers
(of height $V$ and thickness $L$) to go then to results of a more general 
character. 
Let us recall the main features of a {\it time-domain resonance}~\cite{gcv01}.
It follows from the initial condition, given by Eq. \ (\ref{inicon}), that initially there 
is no particle along the tunneling region. Hence evaluating the  probability density  at the 
barrier width $x=L$ as  time evolves from zero, yields a distribution  of 
characteristic times associated to the tunneling process. 
In Fig. \ref{fig1} we plot the  probability density $|\Psi^e|^2$ 
(solid line), normalized to the transmission coefficient $|T_k|^2$, as a 
function of time $t$, corresponding to a potential barrier system with
typical parameters~\cite{mendez}: $V=0.3$ eV, incidence energy 
$E=\hbar^2k^2/2m=0.001$ eV, and effective mass for the electron 
$m=0.067m_e$, for a fixed value of position, $x=L=4.0$ nm. 
We can clearly  appreciate a structure  peaked at the value 
$t_{max}=5.17$ fs. This is  the so called {\it time-domain resonance}.

\begin{figure}[!tbp]
\rotatebox{0}{\includegraphics[width=3.3in]{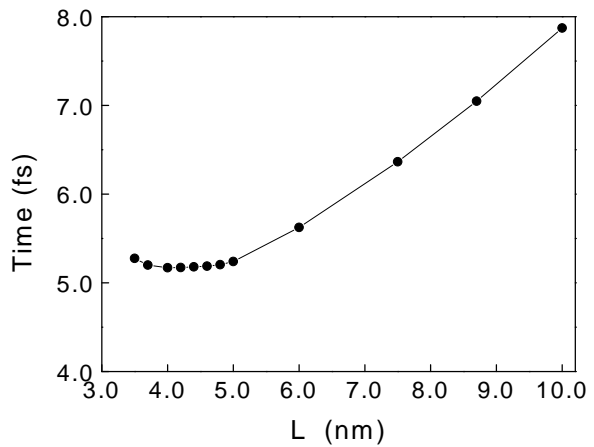}}
\caption{ Maximum of the {\it time-domain resonance} $t_{max}$ (full dot) 
as a function of the barrier width $L$, for an incidence energy $E=.001$ eV.
In this case the barrier height is $V=0.3$ eV. See text.}
\label{fig2}
\end{figure}

In Fig. \ref {fig2} we plot $t_{max}$ (full dot) for the 
same parameters as in the previous figure except for the barrier width $L$, that we vary. 
Here we can clearly observe the existence of a {\it basin} 
along  a range of  values of the barrier width. We can also 
appreciate that if $L$ is further increased, $t_{max}$ starts to grow linearly with $L$. 
Such a linear regime occurs at large barrier widths. We have discussed elsewhere, that 
this last situation refers to non-tunneling processes~\cite{gcv01}.
We have also pointed out in Ref. \cite{gcv01} that for small values of 
the barrier width $L$, the  {\it basin} exhibited 
by $t_{max}$, is the result of a subtle interplay between tunneling
and (non-tunneling) top-barrier resonant processes. 

In what follows we shall investigate under what conditions the time scales 
associated to the {\it basin}, are in fact related to a genuine tunneling process.

\begin{figure}[!tbp]
\rotatebox{0}{\includegraphics[width=3.3in]{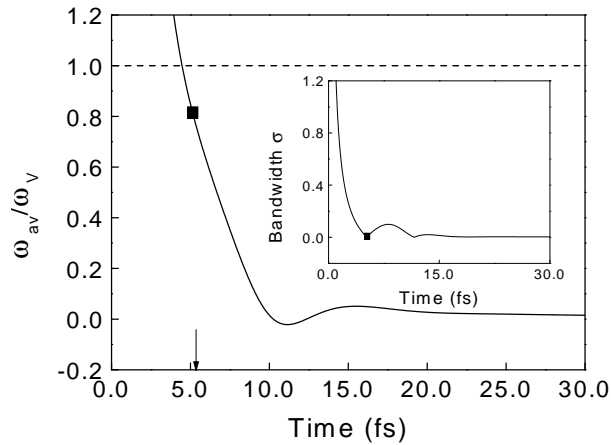}}
\caption{ Relative average local frequency $\omega_{av}/\omega_V$
(solid line) for the case depicted in Fig. \ref{fig1}. 
The cutoff-frequency $\omega_{av}/\omega_V=1$ (dashed line) is included 
for comparison. In the inset we plot the instantaneous bandwidth $\sigma$ 
of the spectrogram depicted in the main graph. Notice that the
frequency deviations at $t_{max}$ are exactly zero, {\it i.e.},
$\sigma(t_{max})=0$. In all cases a full square indicates the 
position of $t_{max}$.}
\label{fig3}
\end{figure}
We use the fact that the initial cutoff wave possesses a distribution of 
momentum components in k-space, and hence also of frequency components. As 
time evolves and the wave interacts with the potential, these frequency 
components manifest themselves in the time evolution of the probability density.
The frequency content of the {\it time-domain resonance} can be investigated  
by performing a time-frequency analysis.
We do this by computing the local average frequency $\omega_{av}$~\cite{mb00,cohen},
\begin{equation}
\omega_{av}=-{\rm Im}\left[\frac{1}{\Psi^s} \frac{d}{dt}\Psi^s
 \right],
\label{4}
\end{equation}
and the instantaneous bandwidth $\sigma$~\cite{cohen}, 

\begin{equation}
\sigma=\left |{\rm Re} \left  [\frac{1}{\Psi^s}\frac{d}{dt}\Psi^s \right ]\right |,
\label{5}
\end{equation}
where $s=i,e$ refers, respectively, to the internal and external solutions.
To exemplify this, we choose the case $L=4.0$ nm depicted in Fig. \ref{fig2},
which is located around the minimum of the {\it basin}. 
In Fig. \ref{fig3} we plot the relative average local frequency 
(relative frequency for short) $\omega_{av}/\omega_V$, where 
$\omega_V=V/\hbar$ is the cutoff frequency,  along the relevant time 
interval, discussed in Fig. \ref{fig1}. We can appreciate that in the 
vicinity of the maximum of the {\it time-domain resonance}, $t_{max}$, 
the probability density is composed  entirely by under-the-barrier
frequency components {\it i.e}  $\omega_{av}/\omega_V<1$. This also occurs 
at the exact value $t_{max}$, also indicated in the figure by a solid square. 
In the inset of Fig. \ref{fig3} we plot the instantaneous bandwidth $\sigma$ of the spectrogram. Notice the absence of a frequency dispersion around the maximum $t_{max}$, {\it i.e.}, $\sigma(t_{max})=0$. The above result indicates that in this case the peak of the {\it time-domain resonance}, and the values close to it, refer to a  tunneling event. 
We have found, however, that this is not a general situation. For instance, for values of $L$ outside the {\it basin}, {\it i.e.}, along the linear regime in Fig.\ {\ref{fig2}), the average frequency related to the corresponding $t_{max}$ is above the cutoff frequency and hence refers to  non-tunneling processes. As we shall present below this is more appropriately
discussed by using the notion of the opacity of the system.
\begin{figure}[!tbp]
\rotatebox{0}{\includegraphics[width=3.3in]{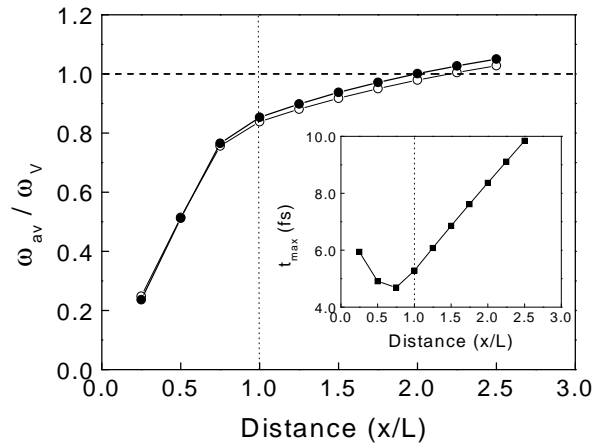}}
\caption{ Relative frequency $\omega_{av}/\omega_V$
of the maximum of the {\it time-domain resonance}, 
as a function of position, measured in units of the barrier width
$L$. The parameters are given in the text. Two incidence energies are considered: 
$E=0.001$ eV (solid dot), and $E=0.01$ eV (hollow dot). In both 
cases, the relative frequency $\omega_{av}/\omega_V$ along the 
internal region is below the cutoff-frequency $\omega_{av}/\omega_V=1$ 
(dashed line). The behavior of $t_{max}$ as a function of position
is illustrated in the inset for the case with $E=0.01$ eV.
The position of the barrier edge, $x=L$, is indicated by 
a dotted line in both figures.}
\label{fig4}
\end{figure}
In Fig. \ref{fig4} we plot the relative frequency $\omega_{av}/\omega_V$
associated to different values of the maximum $t_{max}$, 
measured at different positions along
both the internal and external regions of a potential barrier with
parameters: $V=0.3$ eV, and $L=4.13$ nm.
In this case we choose the following values of the incidence energy: $E=0.001$ eV 
(solid dot), and $E=0.01$ eV (hollow dot). 
In the inset of Fig. \ref{fig4} we show, for the particular case of $E=0.01$ eV, 
the values of $t_{max}$ (solid square) at the different values of position considered
in the main graph.  
As can be clearly appreciated in that figure, the tunneling process along the whole internal region is governed by under-barrier-frequency components, {\it i.e.}, $\omega_{av}/\omega_V<1$. 
We can see in Fig. \ref{fig4}, that we can still observe frequency components below the cutoff-frequency, $\omega_V$ for distances up to $x \simeq 2L$ along the external region.
As $ x/L$ increases further, $ \omega_{av}/\omega_V > 1$.  This behavior 
indicates the prevalence of non-tunneling components in the behavior of the probability 
density~\cite{gcv02}.

We have found that the regime which corresponds to under-the-barrier 
{\it time-domain resonances} at the barrier edge $x=L$,  
may be described more generally by referring to the opacity $\alpha$ 
of the system, defined as,
\begin{equation}
\alpha = \frac{\left [2mV\right]^{1/2}}{\hbar}L,
\label{op}
\end{equation}
and by the dimensionless parameter $u$, the ratio between the potential barrier 
height and the incidence energy, 
\begin{equation}
u=\frac{V}{E}.
\label{u}
\end{equation}
\begin{figure}[!tbp]
\rotatebox{0}{\includegraphics[width=3.3in]{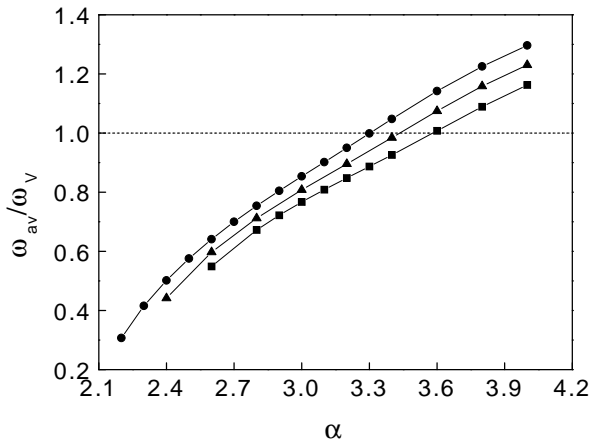}}
\caption{Relative frequency $\omega_{av}/\omega_V$ measured
at the barrier edge $x=L$, as a function of the opacity
$\alpha$. Here we considered a barrier height 
$V=0.3$ eV, and the parameters: $u=300$ (solid dot), $u=10$ 
(solid triangle), and $u=5$ (solid square). Note that for
values of the opacity smaller than $\alpha \simeq 3.3$, the relative
frequencies for all values of $u$ are below the cutoff-frequency
$\omega_{av}/\omega_V=1$ (dashed line). See text.}
\label{fig5}
\end{figure}
To characterize this tunneling regime, we use the fact that 
all systems sharing the same parameters $\alpha$ and $u$
yield the same relative frequency, $\omega_{av}/\omega_V$. This 
regularity arises from a simple rescaling property of the time-
dependent Schr\"{o}dinger's equation and the corresponding initial
condition. By feeding the dimensionless variables $X=x/L$, and
$T=\omega_V t$ in Eqs. (\ref{tdse}) and  (\ref{inicon}), we obtain,   
\begin{equation}
\left[ i \frac{\partial}{\partial T} + \frac{1}{\alpha^2}\frac{\partial ^2}{\partial X^2}%
- 1\right] \chi (X,T)=0,  
\label{tdser}
\end{equation}
with the initial condition
\begin{equation}
\chi (X,T=0) =\left\{ 
\begin{array}{cc}
e^{i\alpha X/\sqrt{u}}-e^{-i\alpha X/\sqrt{u}}, & \quad X \le 0, \\ 
0, & \quad X>0,
\end{array}
\right.   
\label{iniconr}
\end{equation}
where $\chi(X,T)$ is the rescaled time dependent-solution. From Eqs. 
Eqs. (\ref{tdser}) and  (\ref{iniconr}), it is clear that the time-
dependent solution must depend only on the parameters $\alpha$ and $u$. 
That is, for a fixed value of $\alpha$, all the systems with the same 
parameter $u$, yield the same $\chi(X,T)$. 
As a consequence of the above considerations we can write the relative frequency as,
\begin{equation}
\frac{\omega_{av}}{\omega_V}=-{\rm Im}\left[\frac{1}{\chi} \frac{d}{dT}\chi
 \right].
\label{omegarel}
\end{equation}
and use it to characterize the  regime associated with under-the-barrier frequency components. 
In particular, we are interested in defining the range of values of
$\alpha$ where the relative frequencies associated to the {\it time-domain resonance}
are below the cutoff-frequency $\omega_V$. 
In Ref. \cite{gcv02} it is found that  for opacities less than a critical value no {\it time domain resonances} occur.
We denote it by $\alpha_{min}$ and it has the value $\alpha_{min}=2.065$.
In Fig. \ref {fig5} we plot the relative frequency $\omega_{av}/\omega_V$ as 
function of the opacity $\alpha$, for three different values of the 
parameter $u$, $u=5$ (solid dot), $u=10$ (solid triangle), 
and $u=300$ (solid square). In this case we have chosen a value of 
$V=0.3$ eV in the calculation.  
Although in Fig. \ref {fig5} we have considered values of the parameter $u$ 
such that $5 \leq u \leq 300$, the cases corresponding to very large values 
of $u$ ($u\rightarrow \infty$) (not shown here), almost overlap with the case
$u=300$. 
Thus, for very large values of $u$ we find a maximum value for the opacity 
$\alpha_{max} \sim 3.3$. 
Consequently one may define an opacity ``window", in the range of values $2.065 \leq \alpha \leq 3.3$, where 
the relative frequencies are always below the cutoff-frequency $\omega_{av}/\omega_V=1$, 
irrespective of the value of the parameter $u$, namely of the value of the incidence 
energy. Note that the above numerical values refer to the effective mass $m=0.067m_e$ and  clearly will be modified for other values of the effective mass.

It is of relevance to point out that along the {\it basin} region the time scale given by the peak maximum $t_{max}$
differs in a essential way from both the semi-classical B\"uttiker-Landauer and B\"uttiker traversal times, which in addition to exhibit always a linear dependence with $L$, refer to over-the-barrier processes~\cite{mb00,gcv01}. Also 
$t_{max}$ represents a completely different notion than the phase-time, which corresponds to a long-time asymptotic  notion representing a global effect of the potential on the Schr\" odinger's solution, as discussed in Refs.\ ~\cite{gcv01,gcv02}.  

To conclude we  remark that the analytical solution to the time-dependent Schr\"odinger's equation with quantum shutter initial conditions applies in general to arbitrary potentials, provided they vanish beyond a distance, and can also be extended to deal with finite cutoff pulses as discussed in Ref. \cite{gcv02}. The quantum shutter setup provides a consistent procedure to obtain the tunneling time: initially there is no particle along the tunneling region and as time evolves the transient peaked structure exhibited by the probability density at the barrier width provides the relevant time scale for tunneling. This occurs within a range of values of the opacity of the system and is independent of the incidence energy. It is worth noticing that the values of $\alpha$ within the opacity ``window'' may be obtained using typical parameters of semiconductor heterostructures~\cite{mendez}. Also one should stress that at the peak maximum, the {\it time-domain resonance} is governed by a single frequency. That is, the system acts as a frequency filter. To test our results experimentally would require to consider the detection of tunneling particles in time domain at distances close to the interaction region. 

\acknowledgments{The authors thank J. G. Muga for useful discussions and a critical reading of the manuscript, and acknowledge financial support of DGAPA-UNAM under grant No. IN101301.}

\end{document}